\documentclass[12pt]{article}
\usepackage{fullpage,authblk,cite}
\usepackage{amsmath,amsthm,graphicx,amssymb,verbatim,listings}
\usepackage{wasysym,color,enumitem,mathcomp}
\usepackage[T1]{fontenc}     
\usepackage{lmodern, tipa}   
\usepackage[ pdftex, plainpages = false, pdfpagelabels,
                 pdfpagelayout = useoutlines,
                 bookmarks,
                 bookmarksopen = true,
                 bookmarksnumbered = true,
                 breaklinks = true,
                 linktocpage=all,
                 pagebackref=false,
                 colorlinks = true,
                 linkcolor = Blue,
                 urlcolor  = blue,
                 citecolor = BrickRed,
                 anchorcolor = green,
                 hyperindex = true,
                 hyperfigures
                 ]{hyperref}
\usepackage[usenames, dvipsnames]{xcolor}
\usepackage{xifthen} 

\newcommand{\ket}[1]{{\ensuremath{\left| #1 \right\rangle}}}
\newcommand{\bra}[1]{{\ensuremath{\left\langle #1 \right|}}}
\newcommand{\braket}[2]{{\ensuremath{\left\langle #1 \middle| #2
      \right\rangle}}}

\newcommand{\arxiv}[2][]{\ifthenelse{\isempty{#1}}{\href{http://arxiv.org/abs/#2}{{\tt arXiv:\allowbreak{}#2}}} {\href{http://arxiv.org/abs/#2}{{\tt arXiv:\allowbreak{}#2 [#1]}}}}

\newcommand{\booktitle}{\textsl}
\newcommand{\hrefdoi}[2]{\href{https://dx.doi.org/#1}{#2}}

\newcommand{\cH}{\mathcal{H}}

\newcommand{\bbC}{\mathbb{C}}

\begin{document}
\title{On Two Recent Approaches to the Born Rule}
\author[$\dag$]{Blake C.\ Stacey}
\affil[$\dag$]{\small{\href{http://www.physics.umb.edu/Research/QBism}{QBism Group}, Physics Department, University of
    Massachusetts Boston\protect\\ 100 Morrissey Boulevard, Boston MA 02125, USA}}

\date{\small\today}

\maketitle

\begin{abstract}
  I comment briefly on derivations of the Born rule presented by
  Masanes et al.\ and by Hossenfelder.
\end{abstract}

According to the textbooks, the Born rule is the means in quantum theory by
which probabilities are calculated, given a quantum state and a
representation of a measurement. Apart from certain heterodox
individuals~\cite{Fuchs:2013}, the common sentiment has been that
positing the Born rule as a premise of the theory is distasteful, and
various attempts have been made over the years to derive it from
assumptions that are deemed more culturally suitable. I will remark on
two such proposals here.

We will be making some historical comparisons to Gleason's theorem of
1957~\cite{Gleason:1957}. Since I've been surprised more than once by
physicists not knowing about this theorem, a short recapitulation is
in order. Gleason assumes that to each physical system is associated a
Hilbert space, and that each possible measurement upon that system
corresponds to an orthonormal basis of that space. The concept of
probability enters by way of a \emph{frame function,} a map from unit
vectors to the unit interval with the property that the values for
vectors comprising an orthonormal basis always add up
to~1. Importantly, we assume that the probability assigned to a
measurement outcome depends on the vector representing that outcome,
but not on any choice of basis in which that vector might be
embedded. That is, the definition of frame functions makes probability
assignments ``noncontextual'', in the lingo. Gleason's theorem proves
that if the dimension of the Hilbert space $\cH$ is greater than~2,
any frame function $f: \cH \to [0,1]$ must take the form
\begin{equation}
  f(\ket{\phi}) = \bra{\phi}\rho\ket{\phi} \, ,
\end{equation}
where $\rho$ is a positive semidefinite operator of trace~1, or in
other words, a density matrix. So, Gleason's theorem gives the set of
valid states \emph{and} the rule for calculating probabilities given a
state.

It is significantly easier to prove the POVM version of Gleason's
theorem, in which a ``measurement'' is not necessarily an orthonormal
basis, but rather any resolution of the identity into positive
semidefinite operators, $\sum_i E_i = I$. In this case, the result is
that any valid assignment of probabilities to measurement outcomes, or
``effects'', takes the form $p(E) = {\rm tr}(\rho E)$ for some density
operator $\rho$. The math is easier; the conceptual upshot is the
same~\cite{Busch:2003, Caves:2004}.

\section{The State Space of Quantum Mechanics is Redundant}

The first paper I'd like to discuss is ``The measurement postulates of
quantum mechanics are operationally redundant'' by Masanes, Galley and
M\"uller~\cite{Masanes:2019}. The short version of my spiel is that
they present a condition on \emph{states} that seems more naturally to
me like a condition on \emph{measurement outcomes.}  Upon making this
substitution, the Masanes, Galley and M\"uller (MGM) result comes much
closer to resembling Gleason's theorem than they say it does.

I have a sneaky suspicion that a good many other attempted
``derivations of the Born rule'' really amount to little more than
burying Gleason's assumptions under a heap of dubious
justifications. MGM do something more interesting than that. They
start with what they consider the ``standard postulates'' of quantum
mechanics, which in their reckoning are five in number. Then they
discard the last two and replace them with rules of a more qualitative
character. Their central result is that the discarded postulates can
be re-derived from those that were kept, plus the more
qualitative-sounding conditions.

MGM say that the assumptions they keep are about state space, while
the ones they discard are about measurements. But the equations in the
three postulates that they keep could just as well be read as
assumptions about measurements instead. Since they take measurement to
be an operationally primitive notion --- fine by me, anathama to many
physicists! --- this is arguably the better way to go. Then they add a
postulate that has the character of noncontextuality:\ The probability
of an event is independent of how that event is embedded into a
measurement on a larger system. So, they work in the same setting as
Gleason (Hilbert space), invoke postulates of the same nature, and
arrive in the same place. The conclusion, if you take their postulates
about complex vectors as referring to measurement outcomes, is that
``preparations'' are dual to outcomes, and outcomes occur with
probabilities given by the Born rule, thereupon turning into new
preparations.

Let's treat this in a little more detail.

Here is the first postulate of what MGM take to be standard quantum
mechanics:
\begin{quotation}
  \noindent To every physical system there corresponds a complex and
  separable Hilbert space $\mathbb{C}^d$, and the pure states of the
  system are the rays $\psi \in {\rm P}\mathbb{C}^d$.
\end{quotation}
We strike the words ``pure states'' and replace them with ``sharp
effects'' --- an equally undefined term at this point, which can
only gain meaning in combination with other ideas later.

(I spend at least a little of every working day wondering why quantum
mechanics makes use of complex numbers, so this already feels
intensely arbitrary to me, but for now we'll take it as read and press
on.)

MGM define an ``outcome probability function'' as a mapping from rays in
the Hilbert space $\mathbb{C}^d$ to the unit interval $[0,1]$. The
abbreviation OPF is fine, but let's read it instead as \emph{operational
preparation function.} The definition is the same: An OPF is a
function ${\bf f}: {\rm P}\mathbb{C}^d \to [0,1]$. Now, though, it
stands for the probability of obtaining the measurement outcome
$\psi$, for each $\psi$ in the space ${\rm P}\mathbb{C}^d$ of sharp
effects, given the preparation ${\bf f}$. All the properties of OPFs
that they invoke can be justified equally well in this reading. If
${\bf f}(\psi) = 1$, then event $\psi$ has probability 1 of occurring
given the preparation ${\bf f}$. For any two preparations ${\bf f}_1$
and ${\bf f}_2$, we can imagine performing ${\bf f}_1$ with
probability $p$ and ${\bf f}_2$ with probability $1-p$, so the convex
combination $p{\bf f}_1 + (1-p){\bf f}_2$ must be a valid
preparation. And, given two systems, we can imagine that the
preparation of one is ${\bf f}$ while the preparation of the other is
${\bf g}$, so the preparation of the joint system is some composition
${\bf f} \star {\bf g}$. And if measurement outcomes for separate
systems compose according to the tensor product, and this $\star$
product denotes a joint preparation that introduces no correlations,
then we can say that $({\bf f} \star {\bf g})(\psi \otimes \phi) =
{\bf f}(\psi) {\bf g}(\phi)$. Furthermore, we can argue that the
$\star$ product must be associative, ${\bf f} \star ({\bf g} \star
{\bf h}) = ({\bf f} \star {\bf g}) \star {\bf h}$, and everything else
that the composition of OPFs needs to satisfy in order to make the
algebra go.

Ultimately, the same math has to work out, after we swap the words
around, because the final structure is self-dual:\ The same set of rays
${\rm P}\mathbb{C}^d$ provides the extremal elements both of the state
space and of the set of effects. So, if we take the dual of the
starting point, we have to arrive in the same place by the end.

But is either choice of starting point more \emph{natural}?

Beginning with the set of measurement outcomes may help put the
mathematics in conceptual and historical context. For instance, when
proving a no-hidden-variables theorem of the Kochen--Specker type, the
action lies in the choice of measurements, and how the rays that
represent one measurement can interlock with those for
another~\cite{Mermin:1993}. So, from that perspective, putting the
emphasis on the measurements and then deriving the state space is the
more conceptually clean move.

That said, on a deeper level, I don't find either choice all that
compelling. To appreciate why, we need only look again at that arcane
symbol, ${\rm P}\mathbb{C}^d$. That is the setting for the whole
argument, and it is completely opaque. Why the complex numbers? Why
throw away an overall phase? What is the meaning of ``dimension'', and
why does it scale multiplicatively when we compose systems? (A typical
justification for this last point would be that if we have $n$
completely distinct options for the state of one system, and we have
$m$ completely distinct options for the state of a second system, then
we can pick one from each set for a total of $nm$ possibilities. But
what are these options ``completely distinct'' with respect to, if we
have not yet introduced the concept of measurement? Why should
dimension be the quantity that scales in such a nice way, if we have
no reason to care about vectors being orthogonal?) All of this cries
out for a more fundamental understanding~\cite{Stacey:2019}.

\section{Nothing Isn't What It Used To Be}

The second paper I'd like to discuss is by Hossenfelder, originally
titled ``Born's rule from almost
nothing''~\cite{Hossenfelder:2020}. The claim of this paper is nicely
stated in a concise form up front. Let $\ket{\Psi}$ and $\ket{\Phi}$
denote unit-norm elements of a complex vector space $\bbC^N$.
\begin{quotation}
  \noindent \textbf{Claim:}\ The only well-defined and consistent
  distribution for transition probabilities $P_N(\ket{\Psi} \to
  \ket{\Phi})$ on the complex sphere which is continuous, independent
  of~$N$, and invariant under unitary operations is $P_N(\ket{\Psi}
  \to \ket{\Phi}) = |\braket{\Psi}{\Phi}|^2$.
\end{quotation}
Here, ``well-defined'' means that $P_N$ always evaluates to a number
in the unit interval, and ``consistent'' means that if $\{\Phi_i\}$ is
an orthonormal basis, then
\begin{equation}
  \sum_{i=1}^N P_N(\ket{\Psi} \to \ket{\Phi_i}) = 1 \, ,
\end{equation}
and also
\begin{equation}
  P_N(\ket{\Phi_i} \to \ket{\Phi_j}) = \delta_{ij} \, .
\end{equation}

Importantly, we have again an assumption of context-independence,
since the transition probability $P_N$ is posited to be indifferent to
the set in which the final state might be embedded. Whereas in
Gleason's theorem the structure of the possible measurement outcomes was
assumed and the state space derived, here both sets are taken as given
(and indeed identical).

If the initial state is $\ket{\Psi}$ and the possible post-transition
states are $\{\ket{\Phi_i}\}$, then unitary transformations will leave
invariant the 3-vertex Bargmann invariants $\braket{\Psi}{\Phi_i}
\braket{\Phi_i}{\Phi_j} \braket{\Phi_j}{\Psi}$. These will vanish if
the post-transition states are an orthonormal basis, but if we do not
assume Gleason-style context independence, then the Bargmann
invariants can be part of the context and the transition probabilities
can depend upon them. (These invariants can be quite rich
mathematically~\cite{Appleby:2017}, so it is almost a shame that they
don't appear more directly in calculating probabilities!) In that
case, it seems to me, we would have no particular reason to say that
orthogonality should mean zero probability. So, we would have no
particular reason to demand that the post-transition states must form
an orthonormal basis.

Even if we do not wish to incorporate the lovely 3-vertex information,
allowing context dependence opens up the possibilities for
non-Born-rule probability assignments. Let $g$ be any continuous,
nonnegative function with the property that $g(0) = 0$. Then the
generalized Preskill rule~\cite{Fuchs:2011} given by
\begin{equation}
  P_N(\ket{\Psi} \to \ket{\Phi_i}|\{\ket{\Phi_j}\})
  = \frac{g(|\braket{\Psi}{\Phi_i}|)}
  {\sum_j g(|\braket{\Psi}{\Phi_j}|)}
\end{equation}
satisfies the desiderata that $\ket{\Psi}$ transitions to itself with
probability 1 and to an orthogonal $\ket{\Phi_i}$ with probability
0.

Later in the paper, the assumption of continuity is dropped. This
raises further possibilities for contextual, non-Born-rule probability
assignments, even piecewise-continuous ones. For example, suppose that
a measurement corresponds to an orthonormal basis $\{\ket{\Phi_i}\}$,
and let a measurement induce a transition from the initial state
$\ket{\Psi}$ to whichever $\ket{\Phi_i}$ has the largest absolute
overlap $|\braket{\Psi}{\Phi_i}|$. If there is a tie for largest, we
distribute the probability evenly across those outcomes. This rule is
unitarily invariant and independent of the dimension $N$. Moreover, if
one of the $\{\ket{\Phi_i}\}$ is equal to $\ket{\Psi}$, then the state
will remain unchanged with probability 1, and because for any basis at
least one absolute overlap must be greater than zero, a transition
will never occur to an orthogonal state.

\section{Conclusion}

Examining these recent approaches to the Born rule has underlined the
significance of Gleason's noncontextuality assumption. Clearly, it is a
mathematically potent condition. From one perspective, it is
physically rather presumptuous:\ If one takes the traditional view
that going from classical to quantum physics means promoting variables
to Hermitian operators, then why should expectation values not depend
upon \emph{operators,} rather than merely upon eigenvectors taken one
at a time? What conceptual desiderata make a Gleason-style
assumption a \emph{natural} move? Finding the answer, I suspect, will
require beginning before Hilbert spaces and deriving them in turn.

\end{document}